\documentclass[titlepage,twoside,12pt]{article}
\usepackage{amsmath}
\usepackage{amssymb}
\usepackage{amsfonts}
\newcommand{\D}{\displaystyle}

\begin{document}

{\LARGE \bf Two Theories of Special \\ \\ Relativity ?} \\ \\

{\bf Elem\'{e}r E ~Rosinger} \\ \\
{\small \it Department of Mathematics \\ and Applied Mathematics} \\
{\small \it University of Pretoria} \\
{\small \it Pretoria} \\
{\small \it 0002 South Africa} \\
{\small \it eerosinger@hotmail.com} \\ \\

\hfill {\it Dedicated to Marie-Louise Nykamp} \\ \\

{\bf Abstract} \\

Recently, [3], it was shown that Special Relativity is in fact based on one single physical axiom which is that of Reciprocity. Originally, Einstein, [1], established Special Relativity on two physical axioms, namely, the Galilean Relativity and the Constancy of the Speed of Light in inertial reference frames. Soon after, [4,5], it was shown that the Galilean Relativity alone is sufficient for Special Relativity. Here it is important to note that, implicitly, three more assumptions have been used on space-time coordinate transformations, namely, the homogeneity of space-time, the isotropy of space, and a mathematical condition of smoothness type. In [3], a boundedness condition on space-time coordinate transformations is used instead of a usual mathematical smoothness type condition. In this paper it is shown that the respective boundedness condition is closely related to a Principle of Transformation Increment Ratio Limitation, or in short, PTIRL, which has an obvious physical meaning. It is also shown that PTIRL is {\it not} a stronger assumption than that of the mentioned boundedness in [3]. Of interest is the fact that, by formulating PTIRL as a physical axiom, the possibility is opened up for the acceptance, or on the contrary, rejection of this physical axiom PTIRL, thus leading to {\it two} possible theories of Special Relativity. And to add further likelihood to such a possibility, the rejection of PTIRL leads easily to effects which involve unlimited time and/or space intervals, thus are not accessible to usual experimentation for the verification of their validity, or otherwise. \\ \\

{\bf 1. A Most General Setup} \\

As in [3], let $S$ and $S'$ be two reference frames with space-time coordinates $( x, y, z, t )$, respectively, $( x \, ', y \, ', z \, ', t \, ' )$.
Further, let $S'$ move with constant velocity $v$ with respect to $S$, and do so parallel with the $x$-axis in $S$. Lastly, let \\

(1.1)~~~~ $ x = y = z = t = 0 ~\Longrightarrow~ x \, ' = y \, ' = z \, ' = t \, ' = 0 $ \\

which means that at $t = t \, ' = 0$, the origins of coordinates in $S$ and $S'$ coincide. \\

Let us consider the most general possible space-time coordinate transformation, [3], namely, of the form \\

(1.2)~~~~ $ \begin{array}{l}
                               x \, ' = X ( x, y, z, t, v ) \\
                               y \, ' = Y ( x, y, z, t, v ) \\
                               z \, ' = Z ( x, y, z, t, v ) \\
                               t \, ' = T ( x, y, z, t, v )
             \end{array} $ \\ \\

For ease of notation, and following [3], for each $v \in \mathbb{A}_{\cal F}$, let us define a mapping \\

(1.3)~~~~ $ f_v : \mathbb{R}^4 \longrightarrow \mathbb{R}^4 $ \\

where for $x, y, z, t \in \mathbb{R}$, we have \\

(1.4)~~~~ $ \begin{array}{l}
              f_v ( x, y, z, t ) = \\ \\
              ~~~~= ( X ( x, y, z, t, v ), Y ( x, y, z, t, v ), Z ( x, y, z, t, v ), T ( x, y, z, t, v ) )
             \end{array} $ \\ \\

Then (1.2) takes the form \\

(1.5)~~~~ $ u \, ' = f_v ( u  ),~~~~ u, u \, ' \in \mathbb{R}^4,~~ v \in \mathbb{R}$ \\

In these terms, the {\it boundedness} condition in [3] can be formulated as follows \\

(1.6)~~~~ $ \begin{array}{l}
                 \forall~~ v \in \mathbb{R} ~~: \\ \\
                 \exists~~ M > 0 ~~: \\ \\
                 \forall~~ u \in \mathbb{R}^4 ~~: \\ \\
                 ~~~~ || \, u \, || \leq 1 ~~\Longrightarrow~~ || \, f_v ( u ) \, || \leq M
             \end{array} $ \\ \\

where for $a = ( a_1, a_2, a_3, a_4 ) \in \mathbb{R}^4$, we define the norm $|| \, a \, || = | \, a_1 \, | + | \, a_2 \, | + | \, a_3 \, | + | \, a_4 \, |$. \\

The {\it continuity} property used in [3], and which is implied by the boundedness condition (1.6), is as follows \\

(1.7)~~~~ $ \begin{array}{l}
                 \forall~~ v \in \mathbb{R},~~ \epsilon > 0 ~~: \\ \\
                 \exists~~ \delta > 0 ~~: \\ \\
                 \forall~~ u \in \mathbb{R}^4 ~~: \\ \\
                 ~~~~ || \, u \, || \leq \delta ~~\Longrightarrow~~ || \, f_v ( u ) \, || \leq \epsilon
             \end{array} $ \\ \\

However, in [3] it is shown that in view of the {\it homogeneity} of space-time, the transformation (1.5) is {\it additive}, thus the continuity property (1.7) implies the following stronger {\it uniform continuity} \\

(1.8)~~~~ $ \begin{array}{l}
                 \forall~~ v \in \mathbb{R},~~ \epsilon > 0 ~~: \\ \\
                 \exists~~ \delta > 0 ~~: \\ \\
                 \forall~~ u, w \in \mathbb{R}^4 ~~: \\ \\
                 ~~~~ || \, w - u \, || \leq \delta ~~\Longrightarrow~~
                                    || \, f_v ( w ) - f_v ( u ) \, || \leq \epsilon
             \end{array} $ \\ \\

We now {\it introduce} the following Principle of Transformation Increment Ratio Limitation, or in short, PTIRL, in terms of (1.1) - (1.5). Namely, given $v \in \mathbb{R}$ and two sets of space-time coordinates $( x_0, y_0, z_0, t_0), \\ ( x, y, z, t ) \in \mathbb{R}^4$, with their corresponding transformed coordinates $( x \, '_0, y \, '_0, z \, '_0, t \, '_0), ( x \, ', y \, ', z \, ', t \,' ) \in \mathbb{R}^4$ through (1.2), (1.5), then there exist $K, \rho > 0$, such that \\

   $~~~~~~~~~~ || \, P - P_0 \, || \leq \rho ~~\Longrightarrow~~
                                   || \, P \,'  - P \, '_0 \, || \leq K || \, P - P_0 \, || $ \\

where $P_0 = ( x_0, y_0, z_0, t_0), P = ( x, y, z, t ), P \, '_0 = ( x \, '_0, y \, '_0, z \, '_0, t \, '_0), P \, ' = ( x \, ', y \, ', z \, ', t \,' )$. \\

We note that $K$ and $\rho$ may depend on $v$ and $P_0 = ( x_0, y_0, z_0, t_0)$, which are supposed to be given. As for $P \, '_0$ it results from $v$ and $P_0$ through the transformation (1.2), (1.5). Further, $P$ is arbitrary, and then $P \, '$ is given by (1.2), (1.5) applied to $v$ and $P$. \\

Consequently, a more detailed formulation of PTIRL is as follows \\

(PTIRL)~~~~ $ \begin{array}{l}
                      \forall~~ v \in \mathbb{R},~~ P_0 \in \mathbb{R}^4 ~: \\ \\
                      \exists~~ K, \rho > 0 ~: \\ \\
                      \forall~~ P \in \mathbb{R}^4 ~: \\ \\
                      ~~~~ || \, P - P_0 \, || \leq \rho ~~\Longrightarrow~~
                                   || \, P \,'  - P \, '_0 \, || \leq K || \, P - P_0 \, ||
               \end{array} $ \\ \\

This is obviously a {\it local Lipschitz} type continuity property, while (1.8) is a {\it uniform continuity} property. Thus as they stand, none of them is in general stronger, or for that matter, weaker than the other. \\

However, the assumption in [3] is not (1.8), and instead of it, it is (1.6). And clearly, (1.6) leads to the following {\it global uniform Lipschitz} property \\

(1.9)~~~~ $ \begin{array}{l}
                 \forall~~ v \in \mathbb{R} ~~: \\ \\
                 \exists~~ L > 0 ~~: \\ \\
                 \forall~~ u, w \in \mathbb{R}^4 ~~: \\ \\
                 ~~~~ || \, f_v ( w ) - f_v ( u ) \, || \leq L || \, w - u \, ||
             \end{array} $ \\ \\

Indeed, it is an immediate consequence of the additivity of $f_v$ that it is also homogeneous with respect to rational numbers, namely, we have $f_v ( r u ) = r f_v ( u )$, for $v \in \mathbb{R}, u \in \mathbb{R}^4$ and $r \in \mathbb{Q}$. \\
Let therefore $s \in \mathbb{R},~ s > 0$, be such that $r = || \, u \, || + s \in \mathbb{Q}$, then for $w = u / r$, we have $|| \, w \, || < 1$, thus (1.6) gives $|| \, f_v ( w ) \, || \leq M$. However $f_v ( w ) = f_v ( u ) / r$, hence
$|| \, f_v ( u ) \, || \leq M ( || \, u \, || + s )$, and since $s$ can be arbitrary small, we obtain \\

(1.10)~~~~ $ || \, f_v ( u ) \, || \leq M || \, u \, ||,~~~~ v \in \mathbb{R}, u \in \mathbb{R}^4 $ \\

from which (1.9) follows obviously, with $L = M$. \\

We can now conclude that PTIRL is {\it not} a stronger assumption than the boundedness assumption (1.6) in [3], since the latter implies (1.9) which is obviously at least as strong as PTIRL. \\ \\

{\bf 2. Accepting or Rejecting the Physical Axiom PTIRL} \\

In order further to clarify the meaning of PTIRL, let us assume that it does not hold. This means that we have \\

(Non-PTIRL)~~~~ $ \begin{array}{l}
                      \exists~~ v \in \mathbb{R},~~ P_0 \in \mathbb{R}^4 ~: \\ \\
                      \forall~~ K, \rho > 0 ~: \\ \\
                      \exists~~ P \in \mathbb{R}^4 ~: \\ \\
                      ~~~~ || \, P - P_0 \, || \leq \rho \\ \\
                      ~~~~ || \, P \,'  - P \, '_0 \, || > K || \, P - P_0 \, ||
               \end{array} $ \\ \\

These two  inequalities above mean, respectively, that we have \\

(2.1)~~~~ $ | \, x - x_0 \, | + | \, y - y_0 \, | + | \, z - z_0 \, | + | \, t - t_0 \, | \leq \rho $ \\

as well as \\

(2.2)~~~~ $ \begin{array}{l}
                  | \, x \, ' - x \, '_0 \, | + | \, y \, ' - y \, '_0 \, | + | \, z \, ' - z \, '_0 \, | +
                          | \, t \, ' - t \, '_0 \, | > \\ \\
              ~~~~~~~~~~~~~~~~ > K ( \, | \, x - x_0 \, | + | \, y - y_0 \, | +
                          | \, z - z_0 \, | + | \, t - t_0 \, | \, )
             \end{array} $  \\ \\

Clearly, (2.2) means that at least one of the following four relations holds \\

(2.3)~~~~ $ \begin{array}{l}
                  | \, x \, ' - x \, '_0 \, | > K ( \, | \, x - x_0 \, | + | \, y - y_0 \, | +
                          | \, z - z_0 \, | + | \, t - t_0 \, | \, ) / 4 \\ \\
                   | \, y \, ' - y \, '_0 \, | > K ( \, | \, x - x_0 \, | + | \, y - y_0 \, | +
                          | \, z - z_0 \, | + | \, t - t_0 \, | \, ) / 4 \\ \\
                   | \, z \, ' - z \, '_0 \, | > K ( \, | \, x - x_0 \, | + | \, y - y_0 \, | +
                          | \, z - z_0 \, | + | \, t - t_0 \, | \, ) / 4 \\ \\
                   | \, t \, ' - t \, '_0 \, | > K ( \, | \, x - x_0 \, | + | \, y - y_0 \, | +
                          | \, z - z_0 \, | + | \, t - t_0 \, | \, ) / 4
             \end{array} $  \\ \\

Consequently, the negation of PTIRL means the existence of a {\it finite velocity} $v \in \mathbb{R}$ and of a {\it space time event} $P_0 \in \mathbb{R}^4$, such that, no matter how near to $P_0$, there exist space-time events $P$ for which at least one of the ratios between, on one hand, the space-time coordinates of the increments between the transformations of $P$ and $P_0$, and on the other hand, the increment between $P$ and $P_0$, can become {\it arbitrarily large}. \\

Indeed, in view of (2.3), the negation of PTIRL takes the form \\

(Non-PTIRL)~~~~ $ \begin{array}{l}
                      \exists~~ v \in \mathbb{R},~~ P_0 \in \mathbb{R}^4 ~: \\ \\
                      \forall~~ K, \rho > 0 ~: \\ \\
                      \exists~~ P \in \mathbb{R}^4,~~ P \neq P_0 ~: \\ \\
                      ~~~~ | \, x - x_0 \, | + | \, y - y_0 \, | + | \, z - z_0 \, | +
                                                  | \, t - t_0 \, | \leq \rho \\ \\
                      ~~~~~~~~ \mbox{and at least one of the following} \\
                      ~~~~~~~~ \mbox{four relations holds} \\ \\
                      ~~~~ \frac{\D | \, x \, ' - x \, '_0 \, |}{\D \, | \, x - x_0 \, | + | \, y - y_0 \, | +
                                     | \, z - z_0 \, | + | \, t - t_0 \, |} > K \\ \\
                      ~~~~ \frac{\D | \, y \, ' - y \, '_0 \, |}{\D \, | \, x - x_0 \, | + | \, y - y_0 \, | +
                                     | \, z - z_0 \, | + | \, t - t_0 \, |} > K \\ \\
                      ~~~~ \frac{\D | \, z \, ' - z \, '_0 \, |}{\D \, | \, x - x_0 \, | + | \, y - y_0 \, | +
                                     | \, z - z_0 \, | + | \, t - t_0 \, |} > K \\ \\
                      ~~~~ \frac{\D | \, t \, ' - t \, '_0 \, |}{\D \, | \, x - x_0 \, | + | \, y - y_0 \, | +
                                     | \, z - z_0 \, | + | \, t - t_0 \, |} > K
                  \end{array} $ \\ \\ \\

{\bf 3. Two Alternative Theories of Special Relativity ?} \\

The above considerations open up {\it two alternatives} in Special Relativity. Namely, one can - as a {\it physical axiom} - accept PTIRL, that is, the Principle of Transformation Increment Ratio Limitation, or on the contrary, based on certain physical arguments, one can reject it. \\

Now, since one of the essential features of Special Relativity is the {\it limitation} on the velocity of propagation of any physical phenomenon, it appears to be more natural {\it not} to reject PTIRL. \\

However, it is well known, see [3,7] and the literature cited there, that the mentioned velocity limitation is not a perfectly independent axiom of Special Relativity, since it follows from the physical axiom of Galilean Relativity, and in fact, from the physical axiom of Reciprocity, under rather general conditions, as shown in [3]. \\

And here, one can note an interesting fact. \\

Namely, what one adds in [3] to Galilean Relativity, more precisely Reciprocity, in order to obtain the Lorenz Transformations, and thus the relativistic rule of velocity addition, as well as the mentioned velocity limitation, is a {\it boundedness} condition from which a continuity property results, a property closely related to PTIRL, as seen in the sequel. \\
Thus in a way, velocity limitation is assumed, in order to obtain velocity limitation ... \\

This remark is {\it not} a criticism of the approach in [3], and instead, points out the fact that, to a certain extent the Galilean Relativity, or for that matter, Reciprocity - taken all alone and in itself - is {\it not} sufficient in order to obtain Special Relativity. \\

In this regard it may be of interest to recall that Einstein himself kept on numerous occasions presenting Special Relativity as being based on {\it two} physical axioms, namely, the Galilean Relativity {\it and} the velocity limitation, in which the highest possible one is of light in void. \\

We can, therefore, conclude that to the extent one is not rejecting highly discontinuous physical processes, there can be two rather distinct theories of Special Relativity, namely, the usual one, obtainable under PTIRL, for instance, and on the other hand, one that obeys Galilean Relativity, and in fact, merely Reciprocity, and also assumes the homogeneity of space-time and isotropy of space, yet it is described by space-time coordinate transformations more general than the Lorenz ones. \\

In case PTIRL is, however, rejected, then as seen above in (Non-PTIRL), at least one of the four ratios \\

$~~~~~~~~~~ \frac{\D | \, x \, ' - x \, '_0 \, |}{\D \, | \, x - x_0 \, | + | \, y - y_0 \, | +
                                     | \, z - z_0 \, | + | \, t - t_0 \, |} \\ \\ \\
                      ~~~~~~~~~~ \frac{\D | \, y \, ' - y \, '_0 \, |}{\D \, | \, x - x_0 \, | + | \, y - y_0 \, | +
                                     | \, z - z_0 \, | + | \, t - t_0 \, |} \\ \\ \\
                      ~~~~~~~~~~ \frac{\D | \, z \, ' - z \, '_0 \, |}{\D \, | \, x - x_0 \, | + | \, y - y_0 \, | +
                                     | \, z - z_0 \, | + | \, t - t_0 \, |} \\ \\ \\
                      ~~~~~~~~~~ \frac{\D | \, t \, ' - t \, '_0 \, |}{\D \, | \, x - x_0 \, | + | \, y - y_0 \, | +
                                     | \, z - z_0 \, | + | \, t - t_0 \, |} $ \\ \\

becomes arbitrarily large in {\it any} neighbourhood of at least {\it one} space-time event $P_0$. \\

Here however, it is worth noting that the last of the above ratios involves the {\it unboundedness of time}, and not of space, as is the case with the first three ratios. And such time unboundedness is a phenomenon {\it not} so easy to detect under usual experiments. \\

Consequently, an unorthodox kind of Special Relativity, one whose possibility was mentioned above in case PTIRL is rejected, may escape the means of detection by usual experiments which, as a rule, do not involve arbitrarily large time intervals. \\

Needless to say, the detection by usual experiments of the space unboundedness which may result from the first three ratios above is, similarly, not an easy task, in case Non-PTIRL is accepted. \\ \\

{\bf 4. Further Possibilities in Non-Archimedean Space-Times} \\

As seen in [18], the Lorenz transformations of Special Relativity can be obtained within far {\it larger} space-times than the usual four dimensional Minkowski one. Consequently, the above arguments in which the formulation of PTIRL opens up the possibility of two rather different theories of Special Relativity may lead to a yet richer such possibility which will be dealt with elsewhere, since it goes considerably beyond the usual Euclidean framework, thus of that of Minkowski as well, and as such, it requires a considerable preliminary mathematical setup. \\

\end{document}